\documentclass[reprint]{revtex4-1}

\usepackage[T1]{fontenc} 
\usepackage{graphicx}
\usepackage{amsmath}
\usepackage[utf8]{inputenc}
\usepackage{lmodern}
\usepackage{hyperref}				

\begin{document}
\title{Active ideal sedimentation: Exact two-dimensional steady states}
\author{Sophie Hermann}
\email{sophie.hermann@uni-bayreuth.de}
\affiliation{Theoretische Physik II, Physikalisches Institut, Universit\"{a}t Bayreuth, D-$95440$ Bayreuth, Germany}
\author{Matthias Schmidt}
\email{matthias.schmidt@uni-bayreuth.de}
\affiliation{Theoretische Physik II, Physikalisches Institut, Universit\"{a}t Bayreuth, D-$95440$ Bayreuth, Germany}
\date{22 December 2017}

\begin{abstract}
We consider an ideal gas of active Brownian particles that undergo self-propelled
motion and both translational and rotational diffusion under the
influence of gravity. We solve analytically the corresponding
Smoluchowski equation in two space dimensions for steady states. The
resulting one-body density is given as a series, where each term is a
product of an orientation-dependent Mathieu function and a
height-dependent exponential. A lower hard wall is implemented as a
no-flux boundary condition.  Numerical evaluation of the suitably
  truncated analytical solution shows the formation of two different 
  spatial regimes upon increasing Peclet number. These
regimes differ in their mean particle orientation and in their
variation of the orientation-averaged density with height.
\end{abstract}

\maketitle

\section{Introduction}
Active Brownian particles show uncommon behaviour in their collective
motion, aggregation, and motility-induced phase separation
\cite{zoettl}. Due to the self-propulsion, such ``swimmers'' form
prototypical nonequilibrium systems. In experimental setups,
self-propelled particles are realized e.g.\ in the form of bacteria
(such as Escherichia coli) \cite{egleti2} or Janus colloids dispersed
in a suitable solvent \cite{palacci}. Because of the diverse range of
problems and applications, the topic of active matter has received
much current interest \cite{zoettl, egleti2, stark}. In particular
aggregation at confining walls and the influence of gravity were
investigated by theory, simulation and experiment, as we summarize in
the following.

Palacci \textit{et~al.}\ \cite{palacci} made the first step to
experimentally determine nonequilibrium properties of swimmers in
dilute suspensions. The authors developed a special experimental
setup, which allowed them to adjust and stabilize the
$\text{H}_2\text{O}_2$-concentration in the solvent and thereby
regulate the swim velocity of the Janus particles. The mean square
displacement of individual colloidal spheres, which undergo
translational and rotational diffusion, was measured. The experimental
results were found to be in accordance with the theoretical prediction
of Howse \textit{et~al.} \cite{howse}, obtained in the Stokes
regime. In the sedimentation experiment an exponential decay of the
density distribution with increasing height was found, with a
quadratically in swim speed increasing sedimentation length (or
effective temperature).  Under gravity the authors also observed an
accumulation of particles at the bottom of the sample, which could
theoretically be reproduced by Enculescu and Stark~\cite{enculescu}
using classical perturbation theory and numerical simulations.

The three-dimensional numerical results of Ref.~\cite{enculescu},
obtained upon neglecting hydrodynamic and interparticle interactions
due to low swimmer concentration, are in accordance with the
observations of Palacci \textit{et~al.}\ \cite{palacci}. Enculescu and
Stark also discovered a polar order of the particles, at the bottom of
the system directed towards the lower wall, and at the upper region of
the system aligned against the orientation of gravity. According to
their results, such orientational order should increase and
become detectable in experiments, when either the particle radius or
the effective gravitational strength is increased. The results were
based on corresponding Langevin and Fokker-Planck equations for a
dilute suspension, where particle interactions can be neglected.

An analytical solution of the three-dimensional Fokker-Planck equation without gravity in the special case of thin films, and therefore neglecting rotational diffusion, was given by Elgeti and Gompper~\cite{elgeti}. Their solution shows good agreement with their computer simulations obtained via multi-particle collision dynamic of colloids between two near walls. The authors also investigated the limit of small Peclet numbers by expanding the density distribution in spherical harmonics. In both cases, and in their simulations, particle adhesion at boundaries was observed, numerically as an exponential decay for small Peclet numbers and as a power-law decay for larger Peclet numbers,
 analytically as a linear combination of several exponentials.
This aggregation could be explained by the orientation of active particles at walls, here described in second order of Peclet number. As it is a result of the Brownian dynamics of spheres, the authors concluded hydrodynamic interactions would promote the effect of accumulation, since they impede rotational diffusion.

A further analytical solution for a two-dimensional system of active ideal-gas-like Brownian particles was given by Lee~\cite{lee}, who considered an infinite channel without gravity, no interactions except with walls and negligible translational diffusion. In this case the Fokker-Planck equation separates into an angle- and a height-dependent part. The solution consists of a linear combination of exponentially with height decaying Mathieu functions, which characterize the angular dependence of the one-body density. Although Lee discussed the boundary conditions, the constant prefactors have been left undetermined. Due to the further simplification that particles can only move in six different spatial directions in a discretized model, the ratio of particle number at the wall to the number in the channel becomes calculable and the result depends on the strength of rotational diffusion. Space between upper and lower boundary was divided into two regions, each with independent rotational diffusion coefficients, and several cases of two equal as well as of two different coefficients were examined.

Solon \textit{et~al.}\ \cite{solon} compared the behaviour of active Brownian colloids with that of run-and-tumble particles. Both particle types differ in rotation: the former have slow angular diffusion, whereas the latter show discrete jumps in their orientation. Using  fluctuating hydrodynamics the authors investigated the influence of rotation on motility-induced phase separation. 
Furthermore they considered ideal gas-like particles in external fields such as gravity and harmonic traps. Apart from escaping from traps, qualitatively similar results were generated for both models. Calculations of steady state of two-dimensional Brownian swimmers during sedimentation seem to be equivalent to Lee~\cite{lee}, except for an additional gravitational term. Neglecting translational diffusion, the Fokker-Planck equation still separates and analytic solution leads to a sum of products of even Mathieu functions and exponentials. However these authors only took one special summand into account.

One possibility to include boundary conditions was shown by Wagner \textit{et~al.}\ \cite{wagner} for two-dimensional Brownian swimmers in a channel, in a constant flux and in a gravitational field. Based on the separation of the Fokker-Planck equation without translational diffusion, a general density distribution is constructed as a linear combination in accordance with Lee~\cite{lee} and Solon \textit{et~al.}\ \cite{solon}. Even a orientational order of the colloids in the channel as described by Enculescu and Stark~\cite{enculescu} was found. The authors constructed a computational technique to approximate unknown parameters by an iterative fit. The general technique can be transferred to find the expansion factors of further systems. Application to a sedimentation system reproduces previous results in density distribution away from the bottom (e.g.\ as reported in \cite{solon}) and gives an approximate solution near the wall.

In this paper we investigate two-dimensional active Brownian particles
sedimenting in a gravitational field. The interparticle interactions
are neglected. The system is bound by a lower hard wall. We present
the exact analytical solution of the steady state Smoluchowski
equation~(\ref{equ:fokker}) including translational and rotational
diffusion. This case goes beyond the above discussed literature, where
at least one diffusion term was neglected in deriving analytic
solutions. We formulate the considered problem in
section~\ref{chap:problem} and solve it in
section~\ref{chap:solution}. The solution~(\ref{equ:solution})
consists of a series of exponentially in height decaying Mathieu
functions. In either limit of passive particles and of negligible
translation diffusion, known relations are reproduced, such as
e.g.\ the barometric law, see section~\ref{chap:results1}, where we
also derive an asymptotic expansion for the case of small values of
the ratio of swim persistence length and gravitational length.  We
present numerical results in section \ref{chap:results2}. Finally,
conclusions and an outlook are given in
section~\ref{chap:conclusions}.

\section{Active ideal sedimentation} 
\label{chap:problemAndSolution}
\subsection{Formulation of the problem}
\label{chap:problem}
Consider a suspension of active Brownian swimmers with buoyant mass
$m$ sedimenting with velocity $v_\text{g}= m g / \gamma$ due to a
linear gravitational field $- g \textbf{e}_z$, where $g$ is
the gravitational acceleration,
$\gamma$ is the translational friction
constant, and $\textbf{e}_z$ is the unit vector in the (upward)
vertical direction. 
The particle orientation is
described by a unit vector $\boldsymbol{\omega}$, which indicates the
direction of self-propelled motion with constant swim velocity
$s$. This direction changes by continuous rotational diffusion with
diffusion constant $D^{\text{rot}} = k_\text{B} T /
\gamma^{\text{rot}}$, where $T$ indicates the temperature,
$k_\text{B}$ the Boltzmann constant and $\gamma^{\text{rot}}$ the
rotational friction constant. Analogously, the translational diffusion
coefficient is defined as $D = k_\text{B} T / \gamma$.
Due to the small density of colloids in certain experimental solutions, we neglect interactions between the particles in the following, in line with the work by Palacci \textit{et~al.}\ \cite{palacci}, and by Enculescu and Stark~\cite{enculescu}.

The continuity equation~\cite{hansen} relates the one-body density $\rho$, the translational current ${\textbf J}$ and the rotational current ${\textbf J}^\omega$ via
\begin{align}
\frac{\partial}{\partial t}\rho(\textbf{r},\boldsymbol{\omega},t) = - \nabla \cdot \textbf{J}(\textbf{r},\boldsymbol{\omega},t) - \nabla^{\omega} \cdot \textbf{J}^{\omega}(\textbf{r},\boldsymbol{\omega},t),  \label{equ:fokkergen1}
\end{align}
 where $t$ indicates time, $\textbf{r}$ indicates position, $\nabla$ indicates the spatial ($\textbf{r}$) and $\nabla^{\omega}$ the orientational ($\boldsymbol\omega$) derivative. When the combination of swimming and sedimentation is modelled as an external force, $\gamma s \boldsymbol{\omega}-mg\textbf{e}_z$, then in the overdamped limit the translational and rotational currents are, respectively, given by
\begin{align}
\textbf{J}(\textbf{r},\boldsymbol{\omega},t) = &- D \nabla \rho (\textbf{r},\boldsymbol{\omega},t)   \nonumber \\
&+ \left( s \boldsymbol{\omega} - v_\text{g} \textbf{e}_z \right) \rho (\textbf{r},\boldsymbol{\omega},t), \label{equ:fokkergen2}\\
\textbf{J}^{\omega}(\textbf{r},\boldsymbol{\omega},t) = &- D^{\text{rot}} \nabla^{\omega} \rho (\textbf{r},\boldsymbol{\omega},t). \label{equ:fokkergen3}
\end{align}

For steady states the temporal change in the one-particle density
distribution vanishes, $\partial\rho/\partial t=0$.  Assuming further
translational invariance perpendicular to gravity simplifies the
dependence of the one-body density to
$\rho(\textbf{r},\boldsymbol{\omega})=\rho(z,\boldsymbol{\omega})$.  In a
two-dimensional system, the orientation can be described by a single
angle $\theta$, such that $\boldsymbol{\omega}(\theta) =
(\sin\theta,\cos\theta)$, where $\theta=0$ characterises upward facing
particles (in positive $z$-direction) and $\theta = \pm \pi$ indicates
downward facing particles. In combination with the uniform, infinite
$x$-extension of the system, this causes an even (in $\theta$) density
distribution $\rho(z,\theta)=\rho(z, -\theta)$. Furthermore the
orientational Laplace operator simplifies to a second partial
derivative: $(\nabla^{\omega})^2 =\partial^2 / \partial
\theta^2$.\\ Applying these properties and inserting the
currents~(\ref{equ:fokkergen2}) and (\ref{equ:fokkergen3}) in the
continuity equation~(\ref{equ:fokkergen1}) leads to a steady-state
Smoluchowski equation of the form~\cite{enculescu}:
\begin{align}
D \frac{\partial^2}{\partial z^2} \rho(z,\theta) - \left( s \cos \theta  - v_\text{g}\right) \frac{\partial}{\partial z} \rho(z,\theta)\quad &\nonumber \\ 
+ D^{\text{rot}} \frac{\partial^2}{\partial \theta^2} \rho(z,\theta)&=0. \label{equ:fokker}
\end{align}

Due to the $2 \pi-$periodicity of the variable $\theta$, the one-body density is also periodic, $\rho(z, \theta) = \rho(z, \theta \pm 2 \pi)$. As a lower boundary of the system we consider a hard wall at $z=0$. Therefore there is no vertical flux at this height \cite{stark,speck},
\begin{align}
J_z(z=0,\theta) = &- D \left. \frac{\partial}{\partial z} \rho(z,\theta) \right|_{z~=~0} \nonumber \\ 
&+ (s \cos \theta - v_\text{g}) \rho(z=0,\theta) \nonumber \\
=  &\,\, 0.
\label{equ:flux}
\end{align}
Furthermore the  particle density  vanishes for negative values of $z$,
\begin{equation}
\rho(z,\theta)=0 \qquad \forall z<0. \label{equ:densityZ<0}
\end{equation}
With the given boundary and the downward direction of gravity, the density profile also vanishes far away from the wall, 
\begin{align}
\lim \limits_{z \rightarrow \infty} \rho(z,\theta) = 0. \label{equ:lim}
\end{align}

\subsection{Solution of the Smoluchowski equation}
\label{chap:solution}
In order to motivate our
analytic solution of the Smoluchowski equation~(\ref{equ:fokker}), we
first consider three selected special cases.

i) Passively sedimenting particles have no swim velocity, $s=0$, and no preferred orientation, hence $\rho(z,\boldsymbol{\omega}) = \rho(z)$. Their steady state density distribution is described by the celebrated barometric law
\begin{align}
\rho(z) \propto \exp(- v_\text{g} z/ D ) = \exp(- \lambda_\text{eq}z),
\label{equ:solutionp}
\end{align} 
which can be straightforwardly obtained by solving equation~(\ref{equ:fokker}) for the case of $s=0$. Here the gravitational length in equilibrium is 
\begin{align}
\lambda_\text{eq}^{-1} = D / v_\text{g} = k_\text{B} T / (m g). \label{equ:lambdaEq}
\end{align}

ii) For active colloids the solution of the Smoluchowski equation~(\ref{equ:fokker}) with negligible translational diffusion, $D=0$, can be separated into a product of an angle- and a height-dependent part $\rho(z,\theta) \propto f(\theta) p(z)$ \cite{lee, solon, wagner}. It turns out that $p(z) \propto \exp(- \lambda z)$, i.e.\ the result is again exponentially decaying with $\lambda^{-1}$ representing the gravitational length \cite{lee, solon, wagner}.
In this case the constant $\lambda$ is obtained by requiring a $2 \pi$-periodic dependence on $\theta$.

iii) In sedimentation experiments of active colloids one finds an exponentially decaying particle density $\rho(z)$ along the vertical axis: In three dimensions Palacci \textit{et~al.}\ \cite{palacci} found an increasing gravitational length with increasing swim velocity. Ginot \textit{et~al.}\ \cite{ginot} observed the exponential decay in a two-dimensional, dilute suspension of Janus particles on a slightly tilted plane.

Inspired by these special cases and experimental observations, one might expect the same $z$-dependence in the most general case of (\ref{equ:fokker}), despite the fact that a separation of variables is no longer possible. Our chosen ansatz is therefore 
\begin{align}
\rho(z,\theta) \propto f(\theta, \lambda) \exp(- \lambda z), \label{equ:ansatz}
\end{align}
where $\lambda$ is assumed to be constant and both $f$ and $\lambda$ are yet to be determined.
Insertion of the ansatz into (\ref{equ:fokker}) leads to a second order linear differential equation for $f$, which can be reordered as a Mathieu equation 
\cite{Note1}
\begin{align}
\frac{\partial^2}{\partial \eta^2} f(\lambda,\eta) + (a(\lambda) - 2 q(\lambda) \cos(2 \eta)) f(\lambda,\eta) = 0. \label{equ:mathieu}
\end{align} 
Here the angular variable $\theta$ is rescaled as
\begin{align}
 \eta = \theta / 2, \label{equ:par1}
\end{align}
and $a$ and $q$ are independent of $\eta$ and defined as
\begin{align}
a(\lambda) &= 4D \lambda^2 / D^{\text{rot}} - 4 v_\text{g} \lambda / D^{\text{rot}}, \label{equ:par2}\\
q(\lambda) &= -2 s \lambda / D^{\text{rot}}. \label{equ:par3}
\end{align}
The differential equation~(\ref{equ:mathieu}) has the even Mathieu cosine and odd Mathieu sine  functions as solutions. Considering the symmetry of $f(\theta)=f(- \theta)$, only the even Mathieu functions $\text{C}(a,q,\eta)$ are relevant. The corresponding characteristic curve of order $n$, represented by the function $a_n(q(\lambda))$, confines regions in the parameter space of $q$ and $a$ [as given by \eqref{equ:par2} and\eqref{equ:par3}] where the Mathieu functions are stable.
 Since the particle orientation $\theta$ attains values between $-\pi$ and $\pi$, the coordinate transformation \eqref{equ:par1} renders $f(\eta)$ to be $\pi$-periodic in $\eta$. This constitutes a non-trivial condition, because the Mathieu functions are aperiodic in general. Periodicity can be constructed with the special relation 
\begin{align}
  a_n(q(\lambda))=a(\lambda)
  \label{equ:characteristicCurves}
\end{align}
between $a$ and $q$.
For real values of $n$, the Mathieu cosine is periodic with an arbitrary frequency; integer value of $n$ guarantee $2 \pi$-periodicity; even integers $n$ lead to $\pi$-periodicity in $\eta$, as is requested. In order to distinguish from general aperiodic {Mathieu functions $\text{C}(a,q,\eta)$},
 the notation changes to $\text{ce}_n(q,\eta)$ for periodic Mathieu cosine functions.
In order to satisfy the constraint~\eqref{equ:characteristicCurves}, given an even value of $n$, one needs to determine a corresponding suitable value for $\lambda$. This task can be done either by computer algebra systems or by series expansion  \cite{McLachan}.  
 
Assuming all constants of the system (except for $\lambda$) to be positive, there are two solutions for each even order $n$. One solution is positive, $\lambda^+ \ge 0$, and one is negative, $\lambda^- \le 0$. An exception is the case $n=0$ and $v_\text{g} =0$, where the only suitable value is $\lambda=0$.

The complete solution is therefore an infinite linear combination of the solutions for each even $n$. It is hence given as
\begin{align}
\rho(z, \theta) =  \sum \limits_{n = 0}^{\infty} \Big[&b_{2n} \text{ce}_{2n}\left( q( \lambda_{2n}^+), \eta\right) \exp(- \lambda_{2n}^+ z)\nonumber  \\
+ \; &c_{2n} \text{ce}_{2n}\left( q( \lambda_{2n}^-), \eta\right) \exp(- \lambda_{2n}^- z)\Big], \label{equ:solution}
\end{align}
where the remaining free parameters $b_{2n}$ and $c_{2n}$ are constants, which can be determined by the boundary conditions.

The series \eqref{equ:solution} only contains contributions with positive $n$, because the results for positive and negative $n$ are equivalent. We give an alternative derivation of \eqref{equ:solution} in Fourier space in appendix~\ref{chap:fourier}. Reinserting \eqref{equ:solution} into \eqref{equ:fokker} readily proofs that each summand and hence also the linear combination (\ref{equ:solution}) satisfies (\ref{equ:fokker}) by definition of the Mathieu functions. This derivation is shown in detail in appendix~\ref{chap:proof}.

Positive gravitational lengths $( \lambda^+)^{-1}$ belong to $b_{2n}$, so this solution branch can describe normal gravitation. Negative values $\lambda^-$ imply a with $z$ increasing density profile, which can appear for aggregation towards an upper confining wall and for creaming if $m<0$. In the special case of $\lambda=0$ both the Mathieu function and the exponential become constant. This contribution to the solution is hence independent of height and orientation, as is relevant for determining the bulk density in case of no gravity.

Here, for the case of a lower hard wall boundary, a semi-infinite system ($z>0$), an effective particle mass $m>0$, and a gravitational strength $g > 0$, only positive inverse gravitational lengths $\lambda^+$ stay relevant due to the physical limit~(\ref{equ:lim}). Accordingly all factors that belong to any $\lambda^-$ or to $\lambda=0$ are set to zero, $c_{2n} = 0$, which simplifies the one-body density~(\ref{equ:solution}) to
\begin{align}
\rho(z, \theta) = \sum \limits_{n = 0}^{\infty} b_{2n} \text{ce}_{2n}\left( q( \lambda_{2n}), \eta\right) \exp(- \lambda_{2n} z). \label{equ:solutionred}
\end{align} 
The remaining condition of vanishing current~(\ref{equ:flux}) determines the values of $b_{2n}$. A single free scalar parameter remains, which fixes a global prefactor of all constants $b_{2n}$; this can be identified from the integrated density, $\rho_\text{tot}=\iint \mathrm{d}\theta \mathrm{d}z \rho(z,\theta)$ in the system.

Using an alternative ansatz $\rho(z,\theta) \propto f(\theta, \lambda) \exp(\lambda z)$, instead of (\ref{equ:ansatz}), will lead to no additional solutions, because $a_n(q)$ with $n$ an even integer is an even function in $q$ \cite{McLachan}. Therefore only the signs of $\lambda^+$ and $\lambda^-$ interchange, which cancels the different sign of the ansatz and gives the identical results~(\ref{equ:solution}) and (\ref{equ:solutionred}).

\subsection{Limit cases and asymptotic expansion}
\label{chap:results1}
In order to gain a more intuitive understanding of the derived solution (\ref{equ:solutionred}), we consider special limits and its asymptotic behaviour. 

In the case of passive sedimentation (confined again by a hard wall at $z=0$), the particles have no swim velocity, so $s = 0$ and therefore $q=0$, cf.~(\ref{equ:par3}). The Mathieu Cosine in the reduced solution~(\ref{equ:solutionred}) become $ \text{C}(a,q=0,\eta) = \cos \left( \sqrt{a} \eta \right) $, which is a direct consequence of the Mathieu equation. Considering the physical argument that passive colloids have an arbitrary orientation when no torques act, leads us to expect an angle-independent one-body density. The cosine function is constant, when $a(\lambda) =4 \frac{\lambda}{D^{\text{rot}}} \left( D \lambda - v_\text{g} \right)= 0$. Because of the limit~(\ref{equ:lim}), $\lambda =0$ is not a suitable value, so all constants $b_{2n}$ have to vanish apart from the term where $\lambda_\text{eq} = v_\text{g} / D$. The density follows as $\rho(z) \propto \exp( - v_\text{g} z / D)$, i.e.\ the barometric formula (\ref{equ:solutionp}) is recovered correctly.

It is worthwhile to point out that the barometric law is not contained
in the analytic results of Wagner \textit{et~al.}\ \cite{wagner} and
Solon \textit{et~al.}\ \cite{solon}, due to their assumption of
negligible diffusion, $D=0$.  When imposing this condition on our
parameters in the Mathieu function, then \eqref{equ:par3} leaves $q$
unchanged, and \eqref{equ:par2} reduces to $a = - 4 v_\text{g} \lambda
/ D^{\text{rot}}$.  This relation, and considering only the
contribution $n=0$, renders our more general solution
\eqref{equ:solutionred}
 equivalent to the one-body density distribution found by Solon
  \textit{et~al.}\ \cite{solon}.  We hence can interpret their result
  as describing the correct behaviour away from the wall as we show below. 

The complete linear combination for $D=0$ seems consistent with the exponential Mathieu function series of Wagner \textit{et~al.}\ \cite{wagner}, although these authors express their solution with objects, which are only ``related to Mathieu functions'' \cite{wagner}.

We find it useful to define both an active Peclet number $\text{Pe} = s R / D$ and a gravitational Peclet number $\alpha = v_\text{g} R/D$ according to Enculescu \textit{et~al.}\ \cite{enculescu}. 
The particle radius $R$ is obtained from $R^2 = 3D / 4 D^\text{rot}$ and used as a characteristic length scale, e.g.\ in order to define the dimensionless height $\tilde{z}=z/R$. 

To give a more explicit version of $\rho(z,\theta)$ (\ref{equ:solutionred}) and derive further simple, analytic relations, we consider large heights in the system, where the behaviour is hardly influenced by the lower wall.
 Therefore we assume that the order $n=0$ is dominant, as this constitutes the slowest decaying contribution. We demonstrate below in section \ref{chap:results2}, figure \ref{fig:lambda}(a), that this is indeed the case.
Additionally we assume that $|q|$ is small, which corresponds to gravitational lengths  $\lambda_0^{-1}$ larger than the persistence length $s/D^\text{rot}$. 
The expansion of the zeroth order characteristic curve in the limit of small $|q|$ \cite{DLMF},
\begin{align}
a_0(q)= - \frac{1}{2} q^2 + \mathcal{O}(q^4),
\end{align}
can be used to determine $\lambda_0$  (\ref{equ:characteristicCurves}). When taking just the first non-vanishing order into account, then one gets, using (\ref{equ:par2}) and (\ref{equ:par3}), an expression for the asymptotic, nontrivial gravitational length $\hat{\lambda}$ scaled with $\lambda_\text{eq}$ (\ref{equ:lambdaEq}),
\begin{align}
\frac{\lambda_\text{eq}}{\hat{\lambda}} = 1+ \frac{s^2}{2 D D^\text{rot}} = 1 + \frac{2}{3}\text{Pe}^2 = \frac{D_\text{eff}}{D}. \label{equ:Deff}
\end{align}
In the last step $\hat{\lambda}= v_\text{g}/D_\text{eff}$ was assumed, so that (\ref{equ:Deff}) is identical to the value of $D_\text{eff}$ in literature \cite{sevilla, bechinger}, which were often obtained from mean square displacement in long-time limit. 
In three dimensions Palacci \cite{palacci} and Enculescu \cite{enculescu} found a quite similar relation, with the only difference being a $2/9$ instead of the $2/3$ prefactor of $\text{Pe}^2$. 

The limit of small $|q|$ also allows to expand the Mathieu functions
\citep{DLMF} to
\begin{align}
\sqrt{2}\; \text{ce}_0(q, \theta) = 1 - \frac{q}{2} \cos \theta + \frac{q^2}{32} \left( \cos 2 \theta -2 \right) + \mathcal{O}(q^3). \label{equ:approxMathieu}
\end{align}
The entire asymptotic solution $\hat{\rho}$ might be written, based on the results (\ref{equ:Deff}) and (\ref{equ:approxMathieu}) and the more general solution (\ref{equ:solutionred}), as
\begin{align}
\hat{\rho}(z,\theta) = \left( 1 + \frac{s \hat{\lambda}}{D^\text{rot}} \cos \theta \right) \exp(- \hat{\lambda} z).
\end{align}
The corresponding mean polarization of swimmers can be found as 
\begin{align}
\left< \cos \theta \right> = \frac{\iint \mathrm{d} \theta \mathrm{d} z \hat{\rho}(z,\theta) \cos \theta}{\iint \mathrm{d} \theta \mathrm{d} z \hat{\rho}(z,\theta) } = \frac{2 \alpha \text{Pe}}{3 + 2 \text{Pe}^2},
\end{align}
using (\ref{equ:Deff}) and (\ref{equ:approxMathieu}) up to and including linear order in~$q$.
 Again, Enculescu \cite{enculescu} gained a structurally equivalent result for the mean particle orientation, except for a different prefactor, which we attribute to the different spatial dimensionality of both problems.
Note that the polarization of the full system (including all $n$) has to vanish, since there are no acting torques.

\subsection{Numerical results}
\label{chap:results2}
\begin{figure}[b]
\fontsize{9}{10}\selectfont
\includegraphics{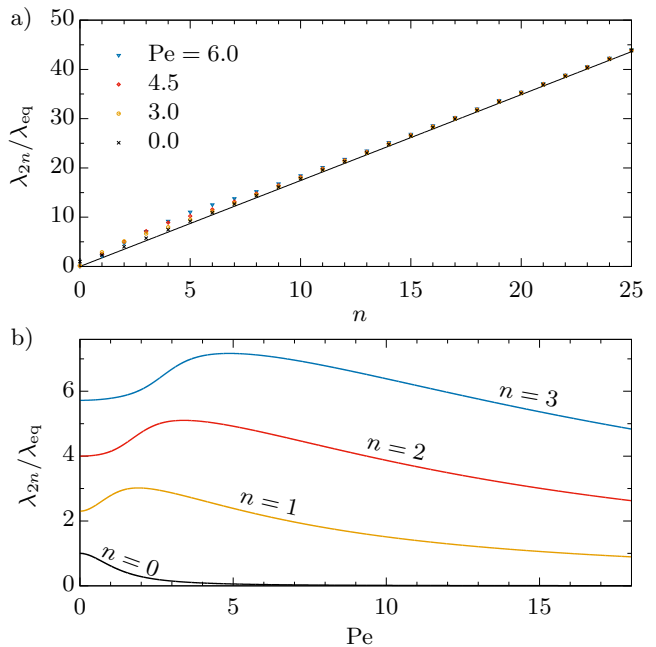}
\\[-2ex]
\caption{\label{fig:lambda} Inverse sedimentation length $\lambda$, cf.~(\ref{equ:characteristicCurves}), scaled with $\lambda_\text{eq}=R / \alpha$. (a) Results are shown as a function of the order $n$ for fixed active Peclet numbers $\text{Pe}=0,3,4.5,6$ (as indicated). The line is a guide to the eye that highlights the asymptotic behaviour. (b) Same as (a) but as a function of $\text{Pe}$ for fixed $n=0,1,2,3$ (as indicated).  In both cases $\alpha = 0.5$ and $R=1$.
}
\end{figure}

In the following we numerically evaluate the full solution \eqref{equ:solutionred} in case of a lower hard wall, \eqref{equ:flux}--\eqref{equ:lim}.
To determine numerical values for the inverse gravitational lengths $\lambda_{2n}$, equation (\ref{equ:characteristicCurves}) is solved numerically (using \textit{Mathematica} 8.0 \cite{Note2}
 and the included normalization of Mathieu functions).
The results, scaled with the equilibrium gravitational length $\lambda_\text{eq}^{-1}$ (\ref{equ:lambdaEq}), are shown in figure \ref{fig:lambda} as a function of the order $n$ (Fig.~\ref{fig:lambda}(a)) and as a function of $\text{Pe}$ (Fig.~\ref{fig:lambda}(b)).

As Fig.~\ref{fig:lambda}(a) demonstrates, a monotonic increase of $\lambda_{2n}$ occurs with increasing value of $n$, approaching a linear behaviour for large orders $n$. 
Changing the value of $\text{Pe}$ has only a small effect on the values of $\lambda_{2n}$; the differences become increasingly small as $n$ grows.
Due to the determined structure of the solution, i.e.\ proportional to a linear combination of $\exp(-\lambda_{2n} z)$, see (\ref{equ:solutionred}),
higher order terms $n$ are important close to the wall, but these decay quickly and hence become less influential for increasing values of $z$. 
 Therefore neglecting all terms except $n=0$ is indeed appropriate far away from the lower wall. 
 Reliable results in regions closer to the wall can be obtained by
including additionally higher orders.

Figure \ref{fig:lambda}(b) shows $\lambda_{2n}/\lambda_\text{eq}$ as a function of $\text{Pe}$. 
Increasing values of $\text{Pe}$ are equivalent
to an increase of the ratio of swim velocity $s$ and a typical
diffusive velocity $D/R$. 
For the case $n=0$ (black curve) a monotonic
decrease of $\lambda_0/\lambda_{\rm eq}$ occurs. The shape is almost identical to the function determined by asymptotic expansion (\ref{equ:Deff}); we have left this curve away in the plot for clarity.
  Hence at large heights we expect an overall expanded distribution $\rho$ (cf.\ (\ref{equ:solutionred})) with raising $\text{Pe}$.
 For $n>0$
an interesting non-monotonic behaviour occurs. While beyond
$\text{Pe}\approx 6$ still a decrease of $\lambda_{
  2n}/\lambda_\text{eq}$ occurs, for smaller values of $\text{Pe}$ an
initial increase of $\lambda_{2n}/\lambda_{\rm eq}$ is found. The
position of the maximum between both types of behaviour shifts to
larger values of $\text{Pe}$ upon increasing the order $n$. We
  expect this type of behaviour to be reflected in the density
  distribution not only in the form of stretching of the orientational averaged density profile $\bar{\rho}$ for large $z$, but also in a compression of $\bar\rho$ for small $z$. This is in accordance with our numerical results discussed below.
 The (further) accumulation in the region of the lower wall arises from those parts of the solution, which are of higher order $n$ and decay fast due to a larger corresponding value of $\lambda_{2n}/\lambda_{\rm eq}$. Regarding the maxima-evolution of $\lambda$ with increasing $n$ this accumulation gets stronger and more concentrated to the value $z=0$ as more active the particles are.
Combining these two aspects of simultaneously expanding and compressing of the one-body density $\rho$, confirms swimming as a mechanism of local particle separation according to orientation, as argued by Elgeti \textit{et~al.}\ \cite{elgeti}.

\begin{figure}[b]
\begin{center}
\fontsize{9}{10}\selectfont
\includegraphics{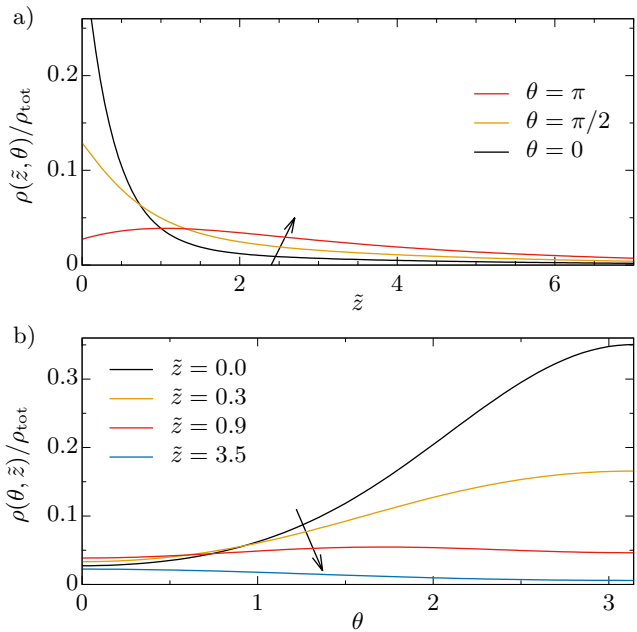} 
\\[-2ex]
\caption{\label{fig:mean_stark} One-body density $\rho(z,\theta)$, normalized by the integrated density $\rho_\text{tot}$. (a) Results are shown  as a function of the dimensionless height $\tilde{z}$ for fixed orientation angle $\theta=0,\pi/2,\pi$ (as indicated by the arrow). (b) Same as (a), but as a function of $\theta$ for fixed $\tilde{z}= 0,0.3,0.9,3.5$ (as indicated by the arrow). In both cases $n=7$, $\text{Pe}=\sqrt{3}$, $\alpha=\sqrt{3/4}$, $R=\sqrt{3/4}$.} 
\end{center}
\end{figure}

In the complete result (\ref{equ:solutionred}), an infinite number of constants $b_{2n}$ still remain to be determined by boundary conditions. 
 In our numerical calculations we include terms up to an order $\tilde{n}$, assuming the distribution $\rho(z,\theta)$ not directly at the wall to be nearly unaffected by higher contributions $n>\tilde n$ (see Fig.~\ref{fig:lambda}). Due to the truncation the boundary condition cannot be satisfied exactly in general. Hence we take the squared flux in $z$-direction at (through) the wall (\ref{equ:flux}) as a cost function. We express the minimization problem for this cost function as a system of linear equations, which we solve numerically for the set of coefficients $b_{2n}$.
For the considered parameter range it is in many cases sufficient to choose $\tilde{n}$ around 10.
 We check adequate convergence by comparing of the average of the density distribution over $z$, $\bar{\rho}(\theta)$, to a constant, which we expect as no external torques are present \cite{elgeti}. Furthermore we monitor the change in $\rho(z, \theta)$ when including one order further, $\tilde{n}+1$, and take care that this change is insignificant on the scale of the plots.

\begin{figure}[b]
\fontsize{9}{10}\selectfont
\includegraphics{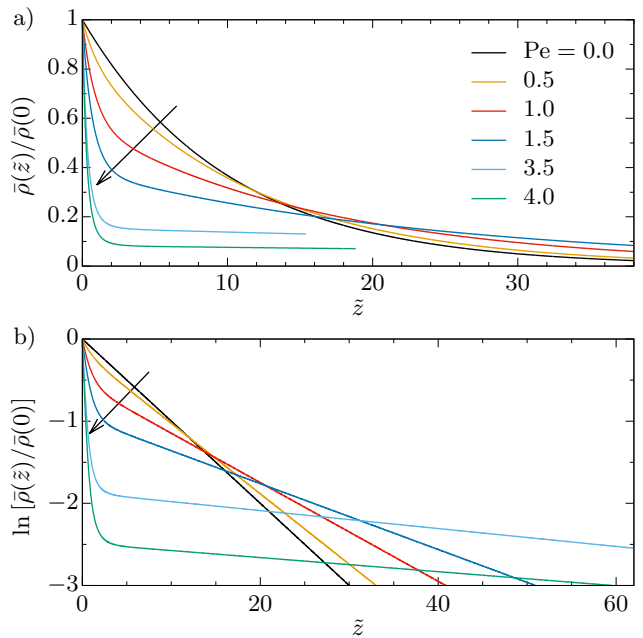}	
\\[-2ex]
\caption{\label{fig:mean_theta} (a) Orientation-averaged one-body density $\bar{\rho}(\tilde{z})$, normalized by its value $\bar{\rho}(\tilde{z}=0)$ at the lower wall,
for different values of $\text{Pe}=0,0.5,1.0,1.5,3.5,4.0$ (as indicated by the arrow). The curves for $\text{Pe} = 3.5, \, 4.5$ are cut off for clarity. (b) Same as (a) but on a logarithmic scale $\bar{\rho}(\tilde{z})/ \bar{\rho}(0)$. The legend is equivalent to (a). The chosen parameter are $\tilde{n}=9$, $\alpha = 0.1$, $R= \sqrt{3/4}$ (except $\tilde{n}=15$ for $\text{Pe} = 3.5$).}
\end{figure}
 
Illustrative, 
 normalized one-body densities $\rho(\tilde{z},\theta)$ as a function  of the dimensionless height $\tilde{z}$ and the angle $\theta$ are shown in figure \ref{fig:mean_stark} (cf.\ \cite{enculescu} for three dimensions). In figure \ref{fig:mean_stark}(a)  different values of orientation are fixed, namely downward, sideward and upward facing swimmers and figure \ref{fig:mean_stark}(b) keeps different selected values of height constant. Downward swimming particles are mostly located near the ground, hindered by the lower wall to swim further down. Upward pointing swimmer are more likely in the bulk of the fluid. This is an aspect, which was also described in literature by Elgeti\ \cite{elgeti} and Enculescu\ \cite{enculescu}, as well as the roughly exponential, with height decaying density $\rho(\tilde{z}, \theta)$ due to gravity (cf.\  \cite{palacci, enculescu, wagner, m.v.}). Gravity is hence the reason for maximum density $\rho$ at the lower wall $\tilde{z}=0$. However, for strong, upward swimming colloids ($\text{Pe}/\alpha > 1$) we find that this maximum position shifts to finite heights $\tilde{z}$, cf.\ Fig.\ \ref{fig:mean_stark}(a).

Altogether the results for two-dimensional one-body densities $\rho$ are qualitative similar to the distributions gained from three dimensional, numerical simulations by Enculescu\ \textit{et al.}\ \cite{enculescu}. 
This accordance extends to the orientational averaged distributions $\bar{\rho}(\tilde{z},\theta)$ with varying active Peclet numbers (Fig. \ref{fig:mean_theta}). 

For comparison, the case of no swimming ($\text{Pe}=0$, $\rho(z)\propto \exp(-\lambda_\text{eq} z)$ according to (\ref{equ:solutionp})) is also shown in figure \ref{fig:mean_theta}(a). With different, finite swim velocity, $\text{Pe}\neq 0$, the curves $\bar{\rho}(z)$ intersect (other parameters unchanged), which is caused by the formation of two regimes. These regimes are even more apparent when plotting the results on a logarithmic-linear scale (Fig. \ref{fig:mean_theta}(b)). Close to the lower wall, one finds a fast decay of $\bar{\rho}(z)$, which changes to a more slowly, single exponential decay, here at approximately $z = 5 R$. As already concluded from figure \ref{fig:lambda}(b), the fast decay gets located more strongly at the lower wall for increasing values of the Peclet number; the difference between the two regimes becomes more apparent.

\section{Conclusion and Outlook} 
\label{chap:conclusions} 
We have presented an analytical solution of the steady state Smoluchowksi equation of sedimenting active Brownian particles in an infinitely dilute suspension confined by a lower hard wall.
We assumed an exponential density decay in height with a constant gravitational length and determined with this ansatz the orientational distribution to be given by Mathieu functions. The linear combination of possible solutions is consistent with two known limits, namely passive colloids in a gravitational field and negligible translational diffusion in the active system. The latter assumption was applied before \cite{lee, solon, wagner} to analytically determine the steady state density distribution of active colloids from the Smoluchowski equation within the approximation.
Since the equilibrium barometric law is not included, one could argue that this is not a good approximation. The solution given in the present work reconciles these limits and it is in accordance with asymptotic cases as well as with qualitative trends of the one-body density distribution, known from experiments (e.g.\ \cite{palacci,ginot}) and simulations (e.g.\ \cite{enculescu}) in three dimensions.

In future work one could try to apply the here presented solution~(\ref{equ:solution}) to further interesting cases. An example might be to investigate the relation of this solution to the analytical result of Elgeti \textit{et~al.}\ \cite{elgeti}. These authors considered active particles without gravity between two closely separated walls with distance $d$, so the inverse rotational Peclet number $D^{\text{rot}} d / s$ could be neglected. They found a strong wall accumulation effect and a one-body density $\rho(z,\theta) \propto \exp( - s d \cos (\theta) z/ D )$, which can be interpreted as containing an angle-dependent inverse gravitational length $\lambda$. It is not clear whether this possibly can be modelled by a series of suitable solutions with constant values of $\lambda$. Furthermore constructing a mathematical proof of the completeness of our solution remains a worthwhile research task for the future.
 
Ginot \textit{et~al.}\ \cite{ginot} observed experimentally an exponential density decay in height far away from the confinement in a two-dimensional sedimentation setup. The gravitational length increased with increasing swim speed of active Janus particles in their dilute suspension. One could try to perform a quantitative comparison between the measured density distribution and the slowest decaying order of our analytical result. It would also be interesting to compare the orientation-dependent density distribution, which would be a worthwhile quantity to be measured in future experiments. 

  Furthermore it would be interesting to take the effect of direct
  interparticle interactions into account, possibly within the power
  functional framework \cite{power,krinninger}. One would then expect
  that packing effects induce correlations on the scale of the
  particles, as occurs in equilibrium \cite{royall}.  Interacting
  active Brownian particles display fluid-fluid phase separation,
  which is known in passive systems to lead to striking phenomena
  under gravity, such as the occurrence of ``floating'' phases
  \cite{floatingLiquid,floatingNematic}. As orientational degrees of
  freedom are relevant in active systems, one might also investigate
  the relationship to sedimentation of passive rotator systems
  \cite{reich,floatingNematic}.  One could attempt to classify the
  occurring phenomena via the concept of sedimentation paths
  \cite{delasheras}, provided that the concept of the chemical
  potential is generalized to nonequilibrium situations
  \cite{rodenburg,dijkstra}.

\appendix
\section{Derivation in Fourier space}
\label{chap:fourier}
In the following an alternative derivation of the series solution~(\ref{equ:solution}) starting from equation~(\ref{equ:fokker}) is shown. The notation is the same as before introduced in section~\ref{chap:problemAndSolution}.
Applying a Fourier transform in the $z$-coordinate to (\ref{equ:fokker}) gives a Mathieu equation
\begin{align}
\frac{\partial^2}{\partial \eta^2} \tilde{\rho}(k,\eta) + (a - 2 q \cos(2 \eta)) \tilde{\rho}(k,\eta) = 0, \label{equ:mathieu0}
\end{align} 
with the constants $a(k) =-4D k^2 / D^{\text{rot}} + 4 i v_\text{g} k / D^{\text{rot}} $, $q(k) =2 i s k/ D^{\text{rot}}$ and 
\begin{align}
\tilde{\rho}(k,\eta) = \int \limits_{-\infty}^{\infty} \mathrm{d}z \exp(- i k z) \rho(z,\eta)
\end{align}
the Fourier transformed, one-body density distribution $\rho(z,\eta)$, where $k$ is the wave number corresponding to $z$.
For the demand of a $2 \pi$-periodic solution in angle $\theta$ the parameter $a$ and $q$ need to satisfy $a_n(q(k))=a(k)$, analogue to equation (\ref{equ:characteristicCurves}), with $n$ an even integer, which constitutes an implicit equation for $k$. However no real solution of $k$ could be found. 
A comparison with passive sedimentation indicates that $k$ might be an imaginary quantity. Those passive particles have a swim velocity of zero and therefore orientation has no influence on their motion. The general Fokker-Planck equation of active colloids (\ref{equ:fokker}) simplifies in this (passive) case to 
\begin{align}
D \frac{\partial^2}{\partial z^2} \rho(z) + v_\text{g} \frac{\partial}{\partial z} \rho(z)=0. 
\label{equ:passive}
\end{align}
Applying the Fourier transform gives
\begin{align}
k \left( -D k - i v_\text{g} \right) \tilde{\rho}(z)=0,
\end{align}
which can be easily solved according to Tailleur and Cates~\cite{cates} by
\begin{align}
\tilde{\rho}(z) = a \delta(k) + b \delta\left(k + i \frac{v_\text{g}}{D} \right)
\end{align} 
Following the reasoning of Tailleur and Cates~\cite{cates} the constant $a$ can be neglected considering the vanishing flux at $z=0$. Thus only imaginary values of $k$ appear in the solution. The inverse transform gives again the well-known barometric formula~(\ref{equ:solutionp}).\\
Transfer of the idea of imaginary wave numbers $k$ to the general case goes along with the ansatz
\begin{align}
\tilde{\rho}(z) \propto f(\theta, k) \delta(k + i \lambda) \label{equ:ansatz2}
\end{align}
containing a Dirac delta function $\delta(\cdot)$.
The function~(\ref{equ:ansatz2}) can be easily inverse transformed back to real space to $\tilde{\rho}(z) \propto f(\theta, - i \lambda) \exp(- \lambda z)$, where $f(\theta, - i \lambda)$ has to satisfy the relation~(\ref{equ:mathieu0}).
It turns out that this condition is the same as equation~(\ref{equ:mathieu}) with equivalent relations for $\theta$,   $a(\lambda)$ and $q(\lambda)$ $\left( \text{cf. } (\ref{equ:par1})-(\ref{equ:par3}) \right)$.
As before, possible values of $\lambda$ are fixed by periodicity in $\theta$ and the linear combination of all suitable expressions results a relation, which is exactly identical to the previous result~(\ref{equ:solution}).

\section{Verification of the solution} \label{chap:proof}
Inserting the solution~(\ref{equ:solution}) on the right hand side of the Fokker-Planck equation~(\ref{equ:fokker}) one gets 
\begin{align}
& D \frac{\partial^2}{\partial z^2} \rho(z,\theta) - \left( s \cos \theta  - v_\text{g}\right) \frac{\partial}{\partial z} \rho(z,\theta) + D^{\text{rot}} \frac{\partial^2}{\partial \theta^2} \rho(z,\theta) \nonumber \\
&= \sum \limits_{n = 0}^{\infty} b_{2n} \Big( D \lambda_{2n}^2 + \lambda_{2n} \left( s \cos \theta  - v_\text{g}\right) 
\notag\\& \qquad\qquad\quad
+ D^{\text{rot}}  \frac{\partial^2}{\partial \theta^2} \Big) \text{ce}_{2n}\left( q, \eta\right)  \exp(- \lambda_{2n} z). \label{equ:proof}
\end{align}
Because of the linear structure of the Smoluchowski equation one could neglect the second solution branch in the verification, $c_{2n} = 0$ $\forall n$, without loss of generality.
Use the definition of the Mathieu equation to determine the second derivative of $\text{ce}_{2n}\left( q( \lambda_{2n}), \eta\right)$ in order to transform (\ref{equ:proof}) to
\begin{align}
& \sum \limits_{n = 0}^{\infty} b_{2n} \Big[ D \lambda_{2n}^2  + \lambda_{2n} \left( s \cos \theta  - v_\text{g}\right) 
\notag\\ &\qquad\qquad
- \frac{D^{\text{rot}}}{4} (a - 2 q \cos\theta )  \Big] \text{ce}_{2n}\left( q, \eta \right) \text{e}^{- \lambda_{2n} z} \nonumber \\
&=\sum \limits_{n = 0}^{\infty} b_{2n} \Big[ \lambda_{2n} \left(D \lambda_{2n} + s \cos \theta  - v_\text{g}\right)
\notag\\&\qquad
- \left(  D \lambda_{2n} -v_\text{g}  - s  \cos\theta \right) \lambda_{2n}  \Big] \text{ce}_{2n}\left( q, \eta\right) \text{e}^{- \lambda_{2n} z} \nonumber \\
&=0.
\end{align}
In the second step, the coefficients $a$ and $q$ were inserted and the summands simplified, which gives zero and is therefore equivalent to the left hand side of equation~(\ref{equ:fokker}).

\providecommand{\noopsort}[1]{}\providecommand{\singleletter}[1]{#1}%

\end{document}